\documentclass[twocolumn,twoside,slac]{revtex4}
\usepackage{graphicx}
\usepackage{fancyhdr}
\usepackage{epsfig}
\usepackage{axodraw}
\usepackage{amsfonts}
\usepackage{amssymb}
\begin{document}

\title{Multivariate Analysis from a Statistical Point of View}


\author{K.S. Cranmer}
\affiliation{University of Wisconsin-Madison, Madison, WI 53706, USA}

\begin{abstract}
Multivariate Analysis is an increasingly common tool in experimental
high energy physics; however, many of the common approaches were
borrowed from other fields.  We clarify what the goal of a multivariate
algorithm should be for the search for a new particle and compare
different approaches.  We also translate the Neyman-Pearson theory
into the language of statistical learning theory.
\end{abstract}

\maketitle

\thispagestyle{fancy}

\section{Introduction\label{S:Intro}}

Multivariate Analysis is an increasingly common tool in experimental
high energy physics; however, most of the common approaches were
borrowed from other fields.  Each of these algorithms were developed
for their own particular task, thus they look quite different at their
core.  It is not obvious that what these different algorithms do
internally is optimal for the the tasks which they perform within high
energy physics.  It is also quite difficult to compare these different
algorithms due to the differences in the formalisms that were used to
derive and/or document them.  In Section~\ref{S:Formalism} we
introduce a formalism for a {\it Learning Machine}, which is
general enough to encompass all of the techniques used within high
energy physics.  In Sections~\ref{S:Searches} \&~\ref{S:NPrisk} we
review the statistical statements relevant to new particle searches
and translate them into the formalism of statistical learning theory.
In the remainder of the note, we look at the main results of
statistical learning theory and their relevance to some of the common
algorithms used within high energy physics.

\section{Formalism}\label{S:Formalism}

Formally a Learning Machine is a family of functions $\mathcal{F}$
with domain $I$ and range $O$ parametrized by $\alpha \in \Lambda$.
The domain can usually be thought of as, or at least embedded in,
$\mathbb{R}^d$ and we generically denote points in the domain as $x$.
The points $x$ can be referred to in many ways ({\it e.g.}  patterns,
events, inputs, examples, \dots).  The range is most commonly
$\mathbb{R}$, $[0,1]$, or just $\{0,1\}$.  Elements of the range are
denoted by $y$ and can be referred to in many ways ({\it e.g.}
classes, target values, outputs, \dots).  The parameters $\alpha$
specify a particular function $f_\alpha \in \mathcal{F}$ and the
structure of $\alpha \in \Lambda$ depends upon the learning
machine~\cite{Vapnik:1968,Vapnik:2000}.

In the modern theory of machine learning, the performance of a
learning machine is usually cast in the more pessimistic setting of
{\it risk}.  In general, the risk, $R$, of a learning machine is written as
\begin{equation}\label{E:risk}
R(\alpha) = \int Q(x,y; \alpha) ~p(x,y) dxdy
\end{equation}
where $Q$ measures some notion of {\it loss} between $f_\alpha(x)$ and
the target value $y$.  For example, when classifying events, the risk
of mis-classification is given by Eq.~\ref{E:risk} with $Q(x,y;\alpha)
= |y - f_\alpha(x)|$.  Similarly, for regression\footnote{During the
presentation, J. Friedman did not distinguish between these two tasks;
however, in a region with $p(x,1) = b$ and $p(x,0)=1-b$, the optimal
$f(x)$ for classification and regression differ.  For classification,
$f(x) = \{1 ~{\rm if} ~b>1/2 , {\rm else} ~0\}$, and for regression the optimal
$f(x)=b$.} tasks one takes $Q(x,y;\alpha) = (y - f_\alpha(x))^2$.
Most of the classic applications of learning machines can be cast into
this formalism; however, searches for new particles place some strain
on the notion of risk.

\section{Searches for New Particles\label{S:Searches}}

The conclusion of an experimental search for a new particle is a
statistical statement -- usually a declaration of discovery or a limit
on the mass of the hypothetical particle.  Thus, the appropriate
notion of performance for a multivariate algorithm used in a search
for a new particle is that performance measure which will maximize the
chance of declaring a discovery or provide the tightest limits on the
hypothetical particle.  In principle, it should be a fairly
straight-forward procedure to use the formal statistical statements to
derive the most appropriate performance measure.  This procedure is
complicated by the fact that experimentalists (and statisticians)
cannot settle on a formalism to use ({\it i.e.} Bayesians {\it vs.}
Frequentists).  As an example, let us consider the Frequentist theory
developed by Neyman and Pearson~\cite{Kendall}.  This was the basis
for the results of the search for the Standard Model Higgs boson at
LEP~\cite{FinalLHWG:2003sz}.

The Neyman-Pearson theory (which we review briefly for completeness)
begins with two Hypotheses: the null hypothesis $H_0$ and the
alternate hypothesis $H_1$~\cite{Kendall}.  In the case of a new
particle search $H_0$ is identified with the currently accepted theory
({\it i.e.} the Standard Model) and is usually referred to as the
``background-only'' hypothesis.  Similarly, $H_1$ is identified with
the theory being tested
%({\it i.e.} Standard
%Model with Higgs boson at some specified mass $m_H$) 
usually referred to as the ``signal-plus-background'' hypothesis%

Next, one defines a region $W\in I$ such that if the data fall in $W$
we accept the null hypothesis (and reject the alternate hypothesis)%
\footnote{With $m$ measurements, we should actually consider the data
as $(x_1,\dots,x_m)\in I^m$, but, for ease of notation, let us only
consider $m=1$.}.  %
Similarly, if the data fall in $I -W$ we reject the null hypothesis
and accept the alternate hypothesis.  
%Recognize that if the null
%hypothesis is true, then there exists a chance that the data could
%fall in $I-W$ and we reject $H_0$ even though it is true -- we commit
%a {\it Type I error}.  
The probability to commit a Type I error is called the {\it size} of
the test and is given by (note alternate use of $\alpha$)
\begin{equation}\label{E:typeI}
\alpha = \int_{I-W} p(x|H_0) dx.
\end{equation}
%Similarly, if the alternate hypothesis is true the data could fall in
%$W$, in which case we accept $H_0$ even though it is false -- we
%commit a {\it Type II error}.
The probability to commit a Type II error is given by
\begin{equation}
\beta = \int_W p(x|H_1) dx.
\end{equation}
Finally, the Neyman-Pearson lemma tells us that the region $W$ of size
$\alpha$ which minimizes the rate of Type~II error (maximizes the
power) is given by
\begin{equation}
W = \left \{x ~ \Bigg | ~\frac{p(x|H_1)}{p(x|H_0)} > k_\alpha \right \}.
\end{equation}

\section{The Neyman-Pearson Theory in the Context of Risk\label{S:NPrisk}}

In Section~\ref{S:Intro} we provided the loss functional $Q$
appropriate for the classification and regression tasks; however, we
did not provide a loss functional for searches for new particles.  
%The first reason for delaying the presentation of the loss functional was
%the lack of consensus within the experimental community on a
%statistical formalism.  
Having chosen the Neyman-Pearson theory as an explicit example, it is
possible to develop a formal notion of risk.

Once the size of the test, $\alpha$, has been agreed upon, the notion
of risk is the probability of Type II error $\beta$.  In order to
return to the formalism outlined in Section~\ref{S:Formalism}, identify
$H_1$ with $y=1$ and $H_0$ with $y=0$.  Let us consider learning
machines that have a range $\mathbb{R}$ which we will compose with a
step function $\tilde{f}(x) = \Theta(f_\alpha(x) - k_\alpha)$ so that
by adjusting $k_\alpha$ we insure that the acceptance region $W$ has
the appropriate size.  The region $W$ is the acceptance region for
$H_0$, thus it corresponds to $W = \{x | \tilde{f}(x) = 0\}$ and $I-W
= \{x | \tilde{f}(x) = 1\}$.  We can also translate the quantities
$p(x|H_0)$ and $p(x|H_1)$ into their learning-theory equivalents
$p(x|0) = p(x,0)/p(0) = \delta(y)p(x,y)/\int p(x,0) dx$ and $\delta(1-y) p(x,y)/
\int p(x,1) dx$, respectively.  With these substitutions we can
rewrite the Neyman-Pearson theory as follows.  A fixed size gives us
the global constraint
\begin{equation}
\alpha = \frac{\int \Theta(f_\alpha(x) - k_\alpha) ~\delta(y)~p(x,y)) dxdy}{\int p(x,0)dx}
\end{equation}
and the risk is given by
\begin{eqnarray}
\beta %&=& \int_W p(x|H_1) dx \\ \nonumber
     &=& \frac{\int  [1-\Theta(f_\alpha(x) - k_\alpha)] ~p(x,1)dx}{\int p(x,1)dx} \\\nonumber
     &\propto & \int \Theta(f_\alpha(x) + k_\alpha) ~\delta(1-y) ~p(x,y)dxdy.
\end{eqnarray}
Extracting the integrand we can write the loss functional as
\begin{equation}\label{E:searchRisk}
Q(x,y;\alpha) = \Theta(f_\alpha(x)+k_\alpha) ~\delta(1-y).
\end{equation}
Unfortunately, Eq.~\ref{E:risk} does not allow for the global
constraint imposed by $k_\alpha$ (which is implicitly a functional of
$f_\alpha$), but this could be accommodated by the methods of Euler
and Lagrange.  Furthermore, the constraint cannot be evaluated without
explicit knowledge of $p(x,y)$.

\subsection{Asymptotic Equivalence}

Certain approaches to multivariate analysis leverage the many powerful
theorems of statistics assuming one can explicitly refer to $p(x,y)$.
This dependence places a great deal of stress on the asymptotic
ability to estimate $p(x,y)$ from a finite set of samples
$\{(x,y)_i\}$.  There are many such techniques for estimating a
multivariate density function $p(x,y)$ given the
samples~\cite{Scott,Cranmer:2000du}.  Unfortunately, for high
dimensional domains, the number of samples needed to enjoy the
asymptotic properties grows very rapidly; this is known as the {\it
curse of dimensionality}.

In the case that there is no (or negligible) interference between the
signal process and the background processes one can avoid the
complications imposed by quantum mechanics and simply add
probabilities.  This is often the case with searches for new
particles, thus the signal-plus-background hypothesis can be rewritten
$p(x,|H_1) = n_s p_s(x) + n_b p_b(x)$, where $n_s$ and $n_b$ are
normalization constants that sum to unity.  This allows us to rewrite
the contours of the likelihood ratio as contours of the
signal-to-background ratio.  In particular the contours of the
%likelihood ratio $\frac{p(x|H_1)}{p(x|H_0)} = k_\alpha$ can be
%rewriten as $\frac{p_s(x)}{p_b(x)} = (k_\alpha - n_b)/n_s =
%k'_\alpha$.
likelihood ratio $p(x|H_1)/p(x|H_0) = k_\alpha$ can be
rewritten as $p_s(x)/p_b(x) = (k_\alpha - n_b)/n_s =
k'_\alpha$.

The kernel estimation techniques described in this conference
represent a particular statistical approach in which classification is
achieved by cutting on a discriminant function $D(x)$~\cite{Askew}.
%The discriminant function $D(x) = \frac{p_s(x)}{p_s(x) + p_b(x)}$ is
The discriminant function $D(x) = p_s(x)/(p_s(x) + p_b(x))$ is
one-to-one with $p_s(x)/p_b(x)$ (which is in turn one-to-one with
the likelihood ratio).  These correspondences are only valid
asymptotically, and the ability to accurately approximate $p(x)$ from
an empirical sample is often far from ideal.  However, for particle
physics applications, up to 5-dimensional multivariate analyses have
shown good performance~\cite{miettinen1}.  Furthermore, they have the
added benefit that they can be easily understood

%\subsection{Priorties in the Non-Asymptotic Regime}
\subsection{Direct vs. Indirect Methods}

The loss functional defined in Eq.~\ref{E:searchRisk} is derived from
a minimization on the rate of Type II error.  This is logically
distinct from, but asymptotically equivalent to, approximating the
likelihood ratio.  In the case of no interference, this is logically
distinct from, but asymptotically equivalent to, approximating the
signal-to-background ratio.  In fact, most multivariate algorithms are
concerned with approximating an auxiliary function that is one-to-one
with the likelihood ratio.  Because the methods are not directly
concerned with minimizing the rate of Type II error, they should be
considered {\it indirect methods}.  Furthermore, the asymptotic
equivalence breaks down in most applications, and the indirect methods
are no longer optimal.  Neural networks, kernel estimation techniques,
and support vector machines all represent indirect solutions to the
search for new particles.  The Genetic Programming (GP) approach
presented in Section~\ref{S:PhysicsGP} is a {\it direct method}
concerned with optimizing a user-defined performance measure.

\section{Statistical Learning Theory}

The starting point for statistical learning theory is to accept that
we might not know $p(x,y)$ in any analytic or numerical form.  This
is, indeed, the case for particle physics, because only $\{(x,y)_i\}$
can be obtained from the Monte Carlo convolution of a well-known
theoretical prediction and complex numerical description of the
detector.  In this case, the learning problem is based entirely on the
training samples $\{(x,y)_i\}$ with $l$ elements.  The risk functional
is thus replaced by the {\it empirical risk functional}
\begin{equation}\label{E:Remp}
R_{\rm emp}(\alpha) = \frac{1}{l} \sum_{i=1}^l Q(x_i, y_i; \alpha).
\end{equation}

There is a surprising result that the true risk $R(\alpha)$ can be
bounded independent of the distribution $p(x,y)$.  In particular, for
$0\le Q(x,y;\alpha) \le 1$
\begin{equation}\label{E:riskLimits}
R(\alpha) {\le} {R_{\rm emp}(\alpha)} 
+ \sqrt{\left(\frac{h(\log(2l/h)+1) - \log(\eta/4)}{l}\right)},
\end{equation}
where $h$ is the Vapnik-Chervonenkis (VC) dimension and $\eta$ is
the probability that the bound is violated.  As $\eta \rightarrow 0$,
$h \rightarrow \infty$, or $l\rightarrow 0$ the bound becomes trivial.

\begin{figure}
\centerline{\epsfig{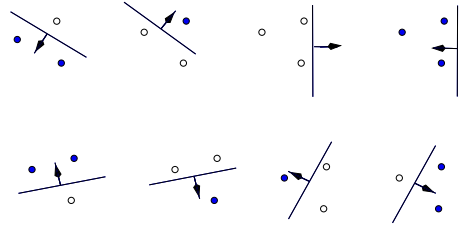}}
\caption{Example of an oriented line shattering 3 points.  Solid and
empty dots represent the two classes for $y$ and each of the $2^3$
permutations are shown.}
\label{fig:shatter}
\end{figure}

The VC dimension is a fundamental property of a learning machine
$\mathcal{F}$, and is defined as the maximal cardinality of a set
which can be shattered by $\mathcal{F}$.  ``A set $\{x_i\}$ can be
shattered by $\mathcal{F}$'' means that for each of the $2^h$ binary
classifications of the points $\{x_i\}$, there exists a $f_\alpha \in
\mathcal{F}$ which satisfies $y_i = f_\alpha(x_i)$.  A set of three
points can be shattered by an oriented line as illustrated in
Figure~\ref{fig:shatter}.  Note that for a learning machine with VC
dimension $h$, not every set of $h$ elements must be shattered by
$\mathcal{F}$, but at least one.

Eq.~\ref{E:riskLimits} is a remarkable result which relates the number
of training examples $l$, a fundamental property of the learning
machine $h$, and the risk $R$ independent of the unknown distribution
$p(x,y)$.  The bounds provided by Eq.~\ref{E:riskLimits} are
relatively weak due to their stunning generality.

It is important to realize that with an independent testing sample one
can evaluate the true risk arbitrarily well.  This testing sample, by
definition, is not known to the algorithm, so the bound is useful for
the design of algorithms through {\it structural risk minimization}.
However, neural networks and most other methods rely on an independent
testing sample to aid in their design and validation.  An independent
testing sample is clearly a better way to assess the true risk of a
multivariate algorithm; however, Eq.~\ref{E:riskLimits} does shed
light on the issues of overtraining, suggests the number of training
samples that are needed, and offers a tool to compare different
algorithms.

\subsection{VC Dimension of Neural Networks}

In order to apply Eq.~\ref{E:riskLimits}, one must determine the
VC dimension of neural networks.  This is a difficult problem in
combinatorics and geometry aided by algebraic techniques.  Eduardo
Sontag has an excellent review of these techniques and shows that the
VC dimension of neural networks can, thus far, only be bounded fairly
weakly~\cite{Sontag}.  In particular, if we define $\rho$ as the
number of weights and biases in the network, then the best bounds are
$\rho^2 < h < \rho^4$.  In a typical particle physics neural network
one can expect $100<\rho<1000$, which translates into a VC dimension
as high as $10^{12}$, which implies $l>10^{13}$ for reasonable bounds
on the risk.  These bounds imply enormous numbers of training samples
when compared to a typical training sample of $10^5$.  Sontag goes on
to show that these shattered sets are incredibly special and that the
set of all shattered sets of cardinality $\mu > 2\rho + 1$ is measure
zero in general.  Thus, perhaps a more relevant notion of the VC
dimension of a neural network is given by $\mu$.

\section{Genetic Programming and Algorithms\label{S:PhysicsGP}}

Genetic Programming (GP) and Genetic Algorithms (GA) are based on a
similar evolutionary metaphor in which ``individuals'' (potential
solutions to the problem at hand) compete with respect to a
user-defined performance measure.  For new particle searches, the rate
of Type II error, the significance, the exclusion potential, or
G. Punzi's suggestion~\cite{Punzi} are all reasonable performance
measures.  Ideally, one would use as a performance measure the same
procedure that will be used to quote the results of the experiment.
For instance, there is no reason (other than speed) that one could not
include discriminating variables and systematic error in the
optimization procedure (in fact, the author has done both).

The use of GP for the classification is fairly limited; however, it
can be traced to the early works on the subject by
Koza~\cite{koza:gp}.
%More recently, Kishore {\it et. al.} extended
%Koza's work to the multicategory problem~\cite{kishore:2000}.  
To the best of the author's knowledge, the first application of GP
within particle physics will appear in~\cite{PhysicsGP}.  The
difference between the algorithms is that GAs evolve a bit string
which typically encodes parameters to a pre-existing program,
function, or class of cuts, while GP directly evolves the programs or
functions.  For example, Field and Kanev \cite{Field:1997kt} used
Genetic Algorithms to optimize the lower- and upper-bounds for six
1-dimensional cuts on Modified Fox-Wolfram ``shape'' variables.  In
that case, the phase-space region was a pre-defined 6-cube and the GA
was simply evolving the parameters for the upper- and lower-bounds.
On the other hand, GP algorithm is not constrained to a pre-defined
shape or parametric form.  Instead, the GP approach is concerned
directly with the construction of an optimal, non-trivial phase space
region ({\it i.e.} an acceptance region $W$) with respect to a
user-defined performance measure.  GPs which only produce polynomial
expressions form a vector space, which allows for a quick
approximation of their VC dimension~\cite{Sontag}.

\section{Conclusions}

Clearly multivariate algorithms will have an increasingly important
role in high energy physics, which necessitates that the field develop
a coherent formalism and carefully consider what it means for a method
to be optimal.  Statistical learning theory offers a formalism that is
general enough to describe all of the common multivariate analysis
techniques, and provides interesting results relating risk, the number
of training samples, and the learning capacity of the algorithm.
However, independent testing samples and the global constraint on the
rate of Type~I error places some strain on the risk formalism.
Finally, when one takes into account limited training data and
systematic errors it is not clear that indirect methods are truly
optimizing an experiments sensitivity.  Direct methods, such as
Genetic Programming, force analysts to be more clear about what
statistical statements they plan to make and remove an artificial
boundary between the goals of the experiment and the optimization
procedures of the algorithm.

\begin{acknowledgments}
This work was supported by a graduate research fellowship from the
National Science Foundation and US Department of Energy Grant
DE-FG0295-ER40896.

\end{acknowledgments}

%\bibliographystyle{apsrev}
%\bibliographystyle{apsrmp}
\bibliographystyle{unsrt}
\bibliography{vbf,bruce_cites,stats,nn,PhysicsGP}

\end{document}